# The Impact of Global Warming on Silicon PV Energy Yield in 2100

Ian Marius Peters, Tonio Buonassisi

Massachusetts Institute of Technology, Cambridge, Massachusetts, 02139, United States of America

*Abstract* — While the installed photovoltaic (PV) capacity grows to a terawatt scale, effects of global climate change unfold. The question arises, how a changing climate, and especially raising temperatures, will affect the performance of PV installations in the future. In this paper we present an estimate of the reduction in energy yield for silicon PV installations due to global warming in the year 2100. Using IPCC global warming scenarios and published temperature coefficients for today's silicon PV panels, we project median reductions in annual energy output of 15kWh/kW$_P$, with reductions up to 50kWh/kW$_P$ in some areas. Higher efficiency cells and advanced cell and module architectures can significantly reduce these losses.

## I. INTRODUCTION

Contemplating the 21st century, two developments seem very likely: 1) the world will get warmer [1], and 2) solar panel deployment will increase [2-4]. Considering these developments in conjunction, the question arises: how will climate change affect solar energy production in the future? It is well known that solar cell performance is affected by local meteorological conditions such as insolation, temperature, water content in the atmosphere [5], and aerosol concentration [6]. The best-established features of current climate change — rising temperature and humidity levels [7] — will result in a performance reduction of all photovoltaic technology, with silicon, the most prevalent photovoltaic technology today [8], being especially sensitive [9].

The impact of global climate change on PV performance has been addressed in a number of studies [10-15]. One challenge that these studies face is the projection of changes in solar insolation over time. An investigation of uncertainties in [10] showed that yield projections that include insolation projections have very large uncertainties. Consequently, there are considerable variations in the projected impact for PV installations for various regions that these studies show. They agree, however, on the overall trend: climate change will likely reduce the power output of solar installations. One parameter that plays a significant role in this reduction is the, arguably, most prominent aspect of climate change: global warming.

In this paper, we present a simple projection of the impact of raising temperatures on the energy yield of silicon PV installations worldwide. The presented results are based on temperature coefficients of PERC type silicon solar cells, and temperature projections according to the RCP4.5 scenario of the Intergovernmental panel on climate change (IPCC).

## II. BACKGROUND

The efficiency of a solar cell is a function of temperature for a variety of reasons. The current-voltage characteristics of a solar cell are governed by Boltzmann's statistic, and the explicit dependence on $k_BT$ results in a reduction of the open-circuit voltage with increasing temperature. There are also device dependent aspects, including the temperature dependence of recombination mechanics [16]. In a first order approximation, the temperature dependence of a PV module can be described by a linear temperature coefficient $T_C$:

$$T_C = \frac{1}{P_M}\frac{d\,P_M}{dT} \quad (1)$$

Where $P_M$ is the power output of the module at the maximum power point [17]. It should be noted that the actual temperature dependence of a solar cell is not necessarily linear over a wide temperature range, but the approximation is often decent for all practically relevant temperatures. In this study we will use a value of -0.45%/K for $T_C$, which can be considered a typical value [18].

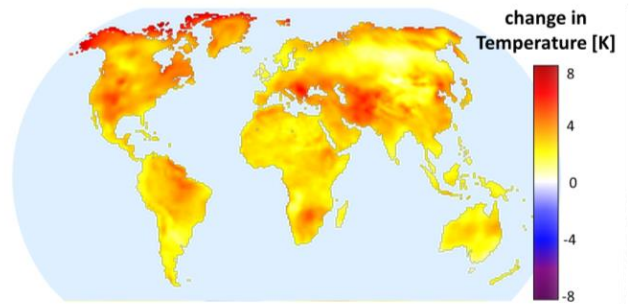

Fig. 1. Projected global temperature change between 2000 and 2100 according to the RCP 4.5 scenario of the IPCC [1]. The average worldwide temperature increase is +1.8K.

Global warming is one of the most well-documented aspects of climate change. Projecting how the temperatures on the planet will evolve is one of the tasks carried out by the IPCC. In this study, we use the Representative Concentration Pathway (RCP) 4.5 scenario to investigate the impact on PV installations. This scenario represents a middle path with greenhouse gas emissions peaking in 2040, and a temperature increase by the end of the century of 1.8K compared to the year 2000. The temperature development around the globe until 2100 is shown in **Figure 1**.

## III. RESULTS

The reduction in energy output is calculated simply by multiplying the worldwide solar resource with the projected temperature change and the temperature coefficient. The result of this calculation is shown in **Figure 2**. As temperatures raise nearly everywhere on the land mass of our planet, energy output is reduced everywhere. Areas that are especially affected include the Southern United States, Southern Africa and Central Asia.

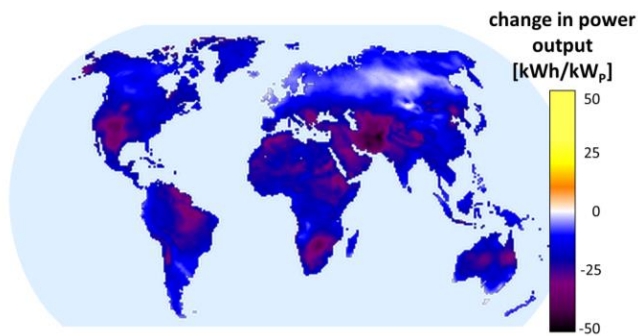

Fig. 2. Projected global change in energy output for a silicon PV installation between 2000 and 2100 for the temperature changes shown in **Figure 1.**

The statistical distribution of the change in energy output over the entire planet is shown in **Figure 3**. The median reduction is approx. 15 kWh/kW$_P$, with some regions reaching up to 50 kWh/kW$_P$.

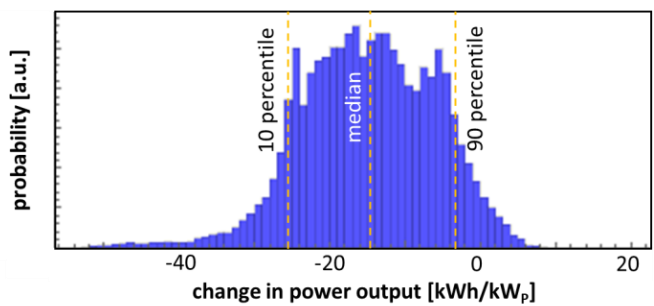

Fig. 3. Histogram of the data shown in **Figure 2**.

## IV. DISCUSSION

The shown results just represent one climate scenario, and use one representative temperature coefficient. Within temperature projections, warming scenarios vary over a wide range, depending on the future development of greenhouse gas emissions. The RCP 4.5 scenarios falls within the lower middle of projected temperature changes. Lower and higher levels are covered by the RCP2.6 and RCP8.5 scenarios, respectively. RCP2.6 assumes a peak in greenhouse gas emissions before the end of the decade, and a drastic reduction in emissions thereafter. The mean likely temperature increase in this scenario is 1.0K. The RCP8.5 scenario assumes raising emissions throughout the rest of the century and project a mean likely temperature increase of 3.7K. Comparing the mean temperature increases in all scenarios gives a good sense of the corresponding projected changes in yield. Hence, in the RCP8.5 scenarios, yield reductions would be more than twice as high as indicated here.

The second factor to affect this calculation is the temperature coefficient of silicon PV modules. -0.45%/K marks a representative value, but coefficients vary depending on technology and cell and module architecture. For silicon PV panels, temperature coefficients between -0.3%/K and 0.6%/K can be found, with the lowest values achieved for HIT type solar cells [19]. Materials with a higher band gap, like CdTe have even lower temperature coefficients [20].

Improvements in cell efficiency have an additional effect on module temperature. As conversion efficiencies increase, more energy is converted into electricity and less into heat [21]. It can be expected, hence, that future solar installations will be more robust towards changes in temperature.

Global warming is not the only relevant aspect of climate change that will affect energy yield. Changes in insolation and humidity will affect the amount of light that reaches a panel. Especially variations in insolation will have an immediate effect that may very well be more noticeable than that of temperature. Insolation variations can largely be considered as an additional factor to the one shown here.

## V. SUMMARY

Global warming has an impact on how photovoltaic installations perform. We use a simple performance model to project the likely reductions in performance of silicon PV panels by the end of the century using climate change scenarios from the IPCC and a representative temperature coefficient. Using the RCP 4.5 scenario and a temperature coefficient of -0.45%/K, we find that the energy output of silicon PV installations will be reduced by approx. 15 kWh/kW$_P$, with some regions experiencing losses of up to 50 kWh/kW$_P$.

ACKNOWLEDGMENT:

Climate scenarios used were from the NEX-GDDP dataset, prepared by the Climate Analytics Group and NASA Ames Research Center using the NASA Earth Exchange, and distributed by the NASA Center for Climate Simulation (NCCS). Insolation data were obtained from the NASA Langley Research Center Atmospheric Science Data Center. This work was financially supported by the DOE-NSF ERF for Quantum Energy and Sustainable Solar Technologies (QESST) and by funding from Singapore's National Research Foundation through the Singapore MIT Alliance for Research and Technology's "Low energy electronic systems (LEES) IRG"

## REFERENCES

[1] IPCC, 2013: Climate Change 2013: The Physical Science Basis. Contribution of Working Group I to the Fifth Assessment Report of the Intergovernmental Panel on Climate Change [Stocker, T.F., D. Qin, G.-K. Plattner, M. Tignor, S.K. Allen, J. Boschung, A. Nauels, Y. Xia, V. Bex and P.M. Midgley (eds.)]. Cambridge University Press, Cambridge, United Kingdom and New York, NY, USA, 1535 pp.

[2] IEA, Market Report Series: Renewables 2017, ISBN 978-92-64-28187-5.

[3] BP Energy Outlook 2018 Edition.

[4] N. Haegel et al., Terawatt-scale photovoltaics: Transform global energy, Science 364:6443 (2019), 836-838.

[5] M. Peters, T. Buonassisi, Energy Yield Limits for Single Junction Solar Cells, Joule 2(6) 2018

[6] M. Peters, S. Karthik, H. Liu, T. Buonassisi, A. Nobre, Urban Haze and Photovoltaics, Energy Environ. Sci., 11 (2018), 3043-3054.

[7] R. Saeger, N. Naik and G. A. Vecchi, Thermodynamic and Dynamic Mecanisms for Large-Scale Changes in the Hydrological Cycle in Response to Global Warming, Journal of Climate, 23 (2010), 4651 – 4668.

[8] Fraunhofer Institute for Solar Energy Systems, ISE, Photovoltaic Status Report, presented Freiburg, 19 June 2018.

[9] O. Dupré, R. Vaillon, M.A. Green, Physics of the temperature coefficients of solar cells, Solar Energy Materials and Solar Cells, 140 (2015), 92 – 100.

[10] Wild M, Folini D, Henschel F, Fischer N, Muller B (2015) Projections of long-term changes in solar radiation based on CMIP5 climate models and their influence on energy yields of photovoltaic systems. Sol Energ 116: 12-24.

[11] J. A. Crook, L. A. Jones, P. M. Forstera and R. Crook, Climate change impacts on future photovoltaic and concentrated solar power energy output, Energy Environ. Sci., 2011, 4, 3101–3109.

[12] M. D. Bartos, M. V. Chester, Impacts of climate change on electric power supply in the Western United States, Nature Climate Change, 5 (2015).

[13] S. Jerez, I. Tobin, R. Vautard, J. P. Montavez, J. M. Lppez-Romero, F. Thais, B. Bartok, O. Bøssing Christensen, A. Colette, M. Deque, G. Nikulin, S. Kotlarski, E. van Meijgaard, C. Teichmann, & M. Wild, The impact of climate change on photovoltaic power generation in Europe, Nature Communications (2015), DOI: 10.1038/ncomms10014.

[14] S. D. Bazyomo, E. A. Lawin, and A. Ouedraogo, Seasonal Trends in Solar Radiation Available at the Earth's Surface and Implication of Future Annual Power Outputs Changes on the Photovoltaic Systems with One and Two Tracking Axes, J Climatol Weather Forecasting 2017, 5:1, DOI: 10.4172/2332-2594.1000201.

[15] V.M. Velasco Herrera, B. Mendoza, G. Velasco Herrera, Reconstruction and prediction of the total solar irradiance: From the Medieval Warm Period to the 21st century, New Astronomy 34 (2015) 221–233.

[16] O. Dupré, R. Vaillon, M.A. Green, Physics of the temperature coefficients of solar cells, Solar Energy Materials and Solar Cells, 140 (2015), 92 – 100.

[17] https://www.pveducation.org/pvcdrom/solar-cell-operation/effect-of-temperature

[18] A. Fell, et al. Input Parameters for the Simulation of Silicon Solar Cells in 2014, IEEE JPV, 5 (2015), 1250 -1263.

[19] J. Lopez-Garcia, D. Pavanello, T. Sample, Analysis of Temperature Coefficients of Bifacial Crystalline Silicon PV Modules, IEEE Journal of Photovoltaics, 8:4 (2018), 960 – 968.

[20] P. K. Dash, N. C. Gupta, Effect of Temperature on Power Output from Different Commercially available Photovoltaic Modules, nt. Journal of Engineering Research and Applications, 5 (2015), 1-4.

[21] Skoplaki, E., and Palyvos, J.A., Operating temperature of photovoltaic modules: a survey of pertinent correlations. Renew. Energy 34 (2009), 23–29.